\documentclass[aps,twocolumn,showpacs,superscriptaddress,prl]{revtex4}

\usepackage{exscale}
\usepackage{graphicx}
\usepackage{amsmath}
\usepackage{latexsym}
\usepackage{amsfonts}
\usepackage{amssymb}
\usepackage{amscd}
\usepackage{bbm}
\usepackage{dcolumn}      
\usepackage{bm}           
\usepackage{pstricks,pst-grad,fancybox,graphics}
\usepackage{enumerate}
\usepackage[sort&compress]{natbib}
\usepackage[latin1]{inputenc} 
\bibpunct{[}{]}{,}{n}{}{}

\newcommand{\bra}[1]{\ensuremath{\langle#1|}}

\newcommand{\ket}[1]{\ensuremath{|#1\rangle}}

\newcommand{\ketbra}[1]{\ensuremath{| #1 \rangle \langle #1 |}}

\newcommand{\eins}{\ensuremath{\mathbbm 1}}
\newcommand{\id}{\ensuremath{\mathbbm 1}}
\newcommand{\HH}{\ensuremath{\mathcal{H}}}

\newcommand{\BE}{\begin{equation}}
\newcommand{\EE}{\end{equation}}
\newcommand{\be}{\begin{equation}}
\newcommand{\ee}{\end{equation}}
\newcommand{\bea}{\begin{eqnarray}}
\newcommand{\eea}{\end{eqnarray}}
\newcommand{\bean}{\begin{eqnarray*}}
\newcommand{\eean}{\end{eqnarray*}}
\newcommand{\kommentar}[1]{}

\newcommand{\mean}[1]{\ensuremath{\langle #1 \rangle}}
\newcommand{\qed}{\ensuremath{\hfill \Box}}

\newcommand{\tr}{{\rm Tr}}

\newcommand{\bc}{\begin{center}}
\newcommand{\ec}{\end{center}}

\renewcommand{\vr}{\ensuremath{\varrho}}

\begin{document}

\title{Covariance matrices and the separability problem}

\author{O.~G\"uhne}
\affiliation{Institut f\"ur Quantenoptik und Quanteninformation,
~\"Osterreichische Akademie der Wissenschaften, 
6020 Innsbruck, Austria,}

\author{P.~Hyllus}
\affiliation{Institut f\"ur Theoretische Physik, Universit\"at
Hannover, Appelstra{\ss}e 2, 30167 Hannover, Germany,}

\affiliation{QOLS, Blackett Laboratory, Imperial College London, 
Prince Consort Road, London SW7 2BW, UK\\ and
Institute for Mathematical Sciences, Imperial College London, 
Prince's Gate, London SW7 2PE, UK}

\author{O.~Gittsovich}
\affiliation{Institut f\"ur Theoretische Physik,
Universit\"at Innsbruck, Technikerstra{\ss}e 25,
6020 Innsbruck, Austria}

\affiliation{Institut f\"ur Quantenoptik und Quanteninformation,
~\"Osterreichische Akademie der Wissenschaften, 
6020 Innsbruck, Austria,}

\author{J.~Eisert}
\affiliation{QOLS, Blackett Laboratory, Imperial College London, 
Prince Consort Road, London SW7 2BW, UK\\ and
Institute for Mathematical Sciences, Imperial College London, 
Prince's Gate, London SW7 2PE, UK}

\date{\today}

\begin{abstract}
We propose a unifying approach to the separability problem 
using covariance matrices of locally measurable observables.
{From} a practical point of view, our approach leads to strong 
entanglement criteria that allow to detect the entanglement of 
many bound entangled states in higher dimensions
and which are at the same time necessary and sufficient 
for two qubits. {From} a fundamental perspective, our 
approach leads to insights into the relations between 
several known entanglement criteria -- such as the computable
cross norm and local uncertainty criteria -- as well as their 
limitations.
\end{abstract}


\maketitle

Entanglement plays a central role in applications of 
quantum information science as well as in the foundations 
of quantum theory. Quite naturally, one of the problems 
that have received a significant amount of attention
is the question to decide whether a given state is 
entangled or separable. In fact, 
the  development of {separability criteria} 
\cite{allsep,ccncrit,loos, algos, algos2, GV} has been one of the key 
activities in quantum information theory: On top of 
{certifying} a given state to be entangled, often in an 
experimental context, they often provide physically valuable 
intuition concerning the structure of the entanglement in 
a given state. 
Among these attempts, methods using positive maps \cite{allsep}
or convex geometry \cite{algos, algos2, GV} turned out to be fruitful.

In this work, we propose a unifying approach to finding 
criteria for separability for finite dimensional systems 
in terms of {covariance matrices} (CMs).
In the infinite-dimensional setting (in 
particular for Gaussian states) such CMs constitute 
a well-established and powerful tool, not least
due to the experimental accessibility of 
quadrature measurements using homodyning
\cite{wernerwolf,vogel,PhilippJens}. In 
contrast, for finite-dimensional systems, 
the theory is yet hardly developed \cite{trotz,OtfriedsPRL}. 
We introduce a  framework 
for CMs for finite-dimensional systems, 
formulate a general separability criterion and
evaluate it for various scenarios. The merits of 
this approach are two-fold:
(i) The  resulting criteria are very strong and allow, notably, 
to detect many bound entangled states. 
(ii) Our approach provides a framework to link 
and understand several 
existing criteria like a recent 
{criterion using the Bloch representation} \cite{vicente},
the one based on {local uncertainty relations} (LURs) \cite{lurs},
or  the {cross-norm} or realignment (CCNR) criterion \cite{ccncrit}, the 
latter being an immediate corollary of our theory.

{\it The main idea. --}
Let us start by defining CMs. 
Let $\vr$ be a given quantum state and let 
$\{ M_k: k=1,\dots ,N \}$ be some observables. Then the 
$N\times N$ CM 
$\gamma $ -- dependent on the state $\vr$ and the 
choice for $\{M_k\}$ -- is given by 
\be
\gamma_{i,j} = 
(\mean{M_i M_j}+ \mean{M_j M_i})/{2}- \mean{M_i} \mean{M_j}.
\ee
This is a real, positive definite 
matrix \cite{robertson} 
and its diagonal entries are just the familiar variances, 
$\gamma_{i,i} = \delta^2(M_i)_\vr$. 
The CM has a useful concavity property: 
if $\vr = \sum_k p_k \vr_k$  
is a convex combination of arbitrary 
states, then 
\be
	\gamma(\vr) \geq \sum_k p_k \gamma(\vr_k).
	\label{concave}
\ee
This 
can be shown to emerge from the generating
function of the moments \cite{OtfriedsPRL,wernerwolf},
reflecting the fact that the variance of an observable 
increases under mixing.

Now, let $\vr$ be a bipartite state on $\HH=\HH_A\otimes \HH_B,$ 
where $d_A$ ($d_B$) is the dimension of $\HH_A$ ($\HH_B$). 
We can choose $d_A^2$ observables $\{A_k\}$ on 
$\HH_A$ such that they form a Hilbert-Schmidt 
orthonormal basis of the observable space, that is, they obey 
$\tr[A_k A_l] = \delta_{k,l}$.
E.g. for a qubit, we may choose normalized Pauli matrices 
including the identity. Similarly, 
we may take $\{B_k\}$ as a basis
of observables in $\HH_B$, and consider the  total set 
$\{M_k\} = \{ A_k \otimes \eins, $ $\eins \otimes B_k\}.$
The CM $\gamma$ then has the block structure
\be
\gamma(\vr,\{M_k\}) =
\begin{bmatrix}
A & C
\\
C^T & B
\end{bmatrix},
\label{gamdef}
\ee
where $A = \gamma (\vr_A, \{A_k\})$ and 
$B = \gamma (\vr_B, \{B_k\})$
are CMs of the reduced states, and $C$ has the entries
$C_{i,j}=\mean{A_i \otimes B_j} - \mean{A_i}\mean{B_j}.$
This matrix will form the starting point to 
characterize the separability properties of 
$\vr$.

To formulate the separability criterion, recall that a state
is {separable} 
iff 
is
a convex combination
of product states, i.e., 
$
\vr = \sum_k p_k \ketbra{a_k, b_k}.
$
We then use Eq.~(\ref{concave}) and the fact that
for a product state the block $C$ in Eq.~(\ref{gamdef})
vanishes, to arrive at the following observation:

\noindent
{\bf Observation 1 (CM criterion).} 
{\it Let $\gamma(\vr)$ be a CM as in 
Eq.~(\ref{gamdef}). If  $\vr$ is separable, then there
exist states $\ketbra{a_k}$ on $\HH_A$, 
$\ketbra{b_k}$ 
on $\HH_B$, 
and convex weights $p_k$ 
such that for 
$\kappa_A = \sum_k p_k \gamma (\ketbra{a_k})$ and
$\kappa_B = \sum_k p_k \gamma (\ketbra{b_k})$ we have \cite{oplusremark}.
\be
	\gamma(\vr,\{M_k\}) \ge \kappa_A \oplus \kappa_B.
	\label{krit1}
\ee
If no such $\kappa_{A/B}$ exist, $\vr$ must be entangled.}

We refer to this criterion as covariance matrix criterion (CMC)
\cite{CVC}. Subsequently, we will show that this
condition can be made an efficient and physically plausible 
test. We will see that as such the criterion is independent
of the choice of the observables $\{A_k\}, \{B_k\}$, but 
that a certification of a violation of the criteria is simplified 
by choosing the Schmidt-basis in operator space or by first using
an appropriate local filtering. 

Let us first note some facts concerning the CM $\gamma$ and the 
$\kappa_{A/B}$, a detailed presentation will be given elsewhere 
\cite{Prep}. A general change of the observables 
$\{M_k\}\mapsto \{\tilde M_k\}$ with 
$\tilde{M}_k = \sum_l \mu_{k,l} M_l$ 
gives rise to a map $\gamma(\varrho,\{\tilde M_k\}) = \mu \gamma (\varrho,\{M_k\}) 
\mu^T$. From this it is easy to see that the CMC is, in principle, 
independent of the choice of the orthonormal $\{A_k\}, \{B_k\}$, 
however, a suitable choice of  them simplifies the falsification 
of Eq.~(\ref{krit1}).
A unitary transformation $\vr \mapsto U \vr U^\dagger$ induces 
a transformation $M_k \mapsto \tilde M_k = U^\dagger M_k U = 
\sum_l O_{k,l} M_l$ where $O$ is orthogonal. The orthogonality of $O$
is equivalent to the statement that the $\{\tilde M_k\}$ are also orthogonal
\cite{loos}. So the eigenvalues of $\gamma(\vr,\{\tilde M_k\})$ 
are invariant under unitary transformations of the state. 
This also means, notably, that by a suitable choice of the 
observables, the matrix $C$ in Eq.~(\ref{gamdef}) can be made 
diagonal by a singular value decomposition.
If $\vr$ is a $d$-dimensional pure state and the $\{M_k\}$ are 
orthogonal, then $\gamma = P/2$ where $P=P^2$ is a 
projector onto a $2(d-1)$-dim subspace of the 
total $d^2$-dim space. This can be directly calculated for a 
specific state and some $\{M_k\}$ 
\cite{kapparemark}. Then, the general 
statement follows 
from the second property. This also implies for Eq.~(\ref{krit1}) that 
$\tr[\kappa_A]= d_A-1$ and  $\tr[\kappa_B]= d_B-1$.

{\it Evaluation for two qubits. --} As a first example, 
we take as observables the normalized Pauli matrices, 
i.e., $\{A_k\} = \{B_k\}= 
\{\eins, \sigma_x , \sigma_y , 
\sigma_z \}/\sqrt{2}$.
Then, in the definition of $\gamma$ in Eq.~(\ref{gamdef}) two 
rows and columns (corresponding to the 
$A_1, B_1 =\eins/\sqrt{2}$) equal zero, thus it suffices 
to consider $\gamma$ as a $6\times 6$ matrix, originating 
only from $\{A_k\}$ and $\{B_k\}$ for $k=2,3,4.$
To characterize the $\kappa_A,$ note that for a pure state
on $\HH_A$ the $3\times 3$ matrix 
$\gamma(\ketbra{a}, \{A_k\})$ 
is a two-dimensional projector, i.e., 
$\gamma = P/2 = (\eins_3-\ketbra{\phi_a} )/2,$ where
$\ket{\phi_a} \in \mathbbm{R}^3.$ In turn,
each matrix of the form $(\eins_3-\ketbra{\phi})/2$ is 
a valid CM, since any such matrix
can be obtained from a special $(\eins_3-\ketbra{\phi_a} )/2$
by an orthogonal transformation (in $\mathbbm{R}^3$),
and for the special case of a qubit there is a one to one 
correspondence between such orthogonal transformations 
and unitary transformations of the state \cite{horos}. 
{From} this, it is easy to see that 
$\kappa_A = (\eins_3 - \rho_A)/2$ where 
$\rho_A$ is a real density matrix on $\mathbbm{R}^3,$ and 
we can summarize: 
\\
{\bf Proposition 2 (Qubit criterion).}
{\it  
Let $\vr$ be a state of two qubits, 
and $\gamma$ be the $6 \times 6$ CM as in Eq.~(\ref{gamdef})
with $\{A_k\} = \{B_k\}= 
\{\sigma_x, \sigma_y,\sigma_z\}/\sqrt{2}$.
$\vr$ fulfills the CMC iff there exist
$3\times 3$ density matrices $\rho_A, \rho_B$ with}
\be
\gamma - \eins_6/2 + (\rho_A \oplus \rho_B)/2 \geq 0. 
\label{qubitlur}
\ee
If we find complex $\rho_{A/B},$ their real part 
saturates Eq.~(\ref{qubitlur}) as well. Finding 
the $\rho_{A/B}$ is a simply solvable 
semi-definite feasibility problem
\cite{algos}. Eq.~(\ref{qubitlur}) will become
important later in the context of LURs.
But let us discuss the general case first.

{\it Evaluation of the CMC  for the general case. --} 
Let us first assume that
$d_A=d_B=d$.

\noindent
{\bf Proposition 3 (General criterion).} 
{\it
Let $\vr$ be a state with $d_A=d_B=d$ and
$A,B,C$ be as in Eq.~(\ref{gamdef}).
If $\vr$ is separable, then}
\bea
2 \sum_{i=1}^{d^2} |C_{i,i}| 
&\leq & 
(\tr[ A] - d + 1)+ 
(\tr[ B] - d + 1)
\nonumber
\\
&=& (1 - \tr[\vr_A^2] ) +  (1 -\tr[\vr_B^2]).
\label{eval1}
\eea

{\it Proof.} First note that a necessary condition for a 
$2 \times 2$ matrix 
$
X=
\begin{bmatrix}
a&c
\\
c&b
\end{bmatrix}
$
to be positive is that $2 |c| \leq a + b.$ If $\vr$ were 
separable, then by the CMC we have 
$ Y= \gamma - \kappa_A \oplus \kappa_B \geq 0.$ This implies that
all $2\times 2$ minor submatrices of $Y$ have to be positive.
Hence, for all $i,j$ 
$2 |C_{i,j}| \leq A_{i,i} + B_{j,j} - (\kappa_A)_{i,i} - (\kappa_B)_{j,j}$.
Summing over all $i,j$ proves the claim.
using $\tr[\kappa_A]= \tr[\kappa_B]= d-1$. 
Eq.~(\ref{eval1}) always holds since
$\sum_k A_k^2 = d \eins$ \cite{ccn}, hence
$\sum_i A_{i,i} = \sum_k \delta(A_k)^2 = d - \tr[\vr_A^2]. $
\qed

As explained before, one may choose $C$ diagonal, 
so considering the diagonal only is no restriction 
of generality \cite{chinese}. One may further try 
to improve Proposition 3 by taking $4\times4$ submatrices 
into account \cite{Prep}. 
Physically, Proposition 3 states that if the correlations  
$C_{i,i}$ are large, then $\vr$ is entangled. 
The question remains how to find those observables for 
which the $C_{i,i}$ are large.
This problem can be overcome by making use of the 
{\it Schmidt decomposition in operator space} \cite{ccn}. 
Recall that we can express any state as
$
	\vr = \sum_k \lambda_k ( G_k^A \otimes G_k^B).
$
Here, $\lambda_k \geq 0$ for all $k$, 
and the $\{G_k\}$ form an orthonormal basis of the operator 
space. Denoting $g_k^{A/B} =\tr[G_k^{A/B}]$ 
we can now use the set $\{G_k^{A/B}\}$ in Proposition 3:

\noindent
{\bf Proposition 4 (CMC for Schmidt form states).} 
{\it If $\vr$ is separable, then}
\begin{equation*}
	2 \sum_k |\lambda_k - \lambda_k^2 g_k^A g_k^B | 
	\leq 2 - \sum_k \lambda_k^2 \bigl((g_k^A)^2 +(g_k^B)^2\bigr).
\end{equation*}
This is a direct application of  Proposition 3 to the diagonal 
of $C.$ One can find examples of states, which are detected by 
the CMC via Proposition 2, but not by Proposition 4 \cite{numerics}. 
However, Proposition 4 connects now the CMC to the CCNR criterion 
\cite{ccncrit,ccn}:

\noindent
{\bf Corollary 5 (Connection to the CCNR criterion).} 
{\it If $\vr$ is separable, then $\sum_k \lambda_k \leq 1$.}

This follows from Proposition 4 and the general 
relation $a^2 +b^2 \geq 2ab.$ Since the condition
$\sum_k \lambda_k \leq 1$ is just the CCNR criterion, 
this implies that the well-known CCNR criterion is a 
direct corollary of our theory \cite{Remarkable}.
 
{\it Enhancing the criterion via local filtering. --} 
A general strategy strengthening the presented test
is the joint use of the CMC with {\it local filtering 
operations}. 
Such operations are maps of the form
$\vr \mapsto \tilde\vr = (F_A \otimes F_B) \vr  (F_A \otimes F_B)^\dagger$
where the $F_{A/B}$ are arbitrary invertible matrices. They preserve the 
entanglement or separability of a given state. 
Local filtering transforms a generic (full rank) $\vr$ into
\begin{eqnarray}
	\tilde \vr 
	=\frac{1}{d_A d_B}
	\biggl(\id + 
	\sum_{i=1}^{d_A^2-1} \xi_i (\tilde{G}_i^A \otimes \tilde{G}_i^B)
	\biggr)
	\label{nnf}
\end{eqnarray}
where  $\xi_i \geq 0$ for all $i$ and the $\{\tilde{G}_i^{A/B}\}$ 
are traceless and orthogonal observables \cite{frank, OldHoro, norway}. 
The matrices $F_A, F_B$ can be found constructively \cite{remark}. We 
will refer to this form as the {filter normal form (FNF)}.
The extraordinarily helpful property is that for a state in the FNF with the 
$\{\tilde{G}_i^{A/B}\}$ as observables the blocks in the CM are diagonal.
We have then in Eq.~(\ref{gamdef})
$A={\rm diag}(0,1,1,\dots,1) / d_A$,
$B={\rm diag}(0,1,1,\dots,1) / d_B$,
and
$C={\rm diag}(0,\xi_1,\xi_2,\dots,\xi_{d_A^2-1})/(d_A d_B),$ 
and obtain:

\noindent
{\bf Proposition 6 (CMC under filtering).} 
{\it
If a generic state $\vr$ is separable and 
$d_A=d_B=d$ we have in its FNF}
        $\sum_{i=1}^{d^2-1} 
        \xi_i \leq d^2- d$.

This is a very strong criterion for separability, 
as some examples will show below. Interestingly, it is 
also necessary and sufficient for two qubits \cite{rankremark}: 
For them, it is easy to see that the $\{\tilde{G}_i^{A/B}\}$ 
in Eq.~(\ref{nnf}) are effectively the Pauli matrices, 
and the separable states of this form are known to be an 
octahedron inside some tetrahedron \cite{horos}, the borders 
of which are described by Proposition 6. Alternatively, 
this can be seen from the results in Ref.~\cite{norway}.

Let us consider the asymmetric case, when $d_A < d_B.$ 
The CM $\gamma$ in the FNF is then similar  as before, 
however $A$ and $B$ are not of the same dimension. 
Two observations are now helpful:
First, we do not sum over all 
$B_{i,i}.$ On the other hand, we cannot subtract all of the 
$(\kappa_{B})_{i,i}$ 
anymore, since $d_B^2-d_A^2$ diagonal elements
of $\kappa_B$ do not occur in the sum.
Moreover, $\gamma$ 
consists of a linear part $\gamma^L_{i,j}= \mean{M_i M_j + M_j M_i}/2$
and a nonlinear part $\gamma^N_{i,j}=\mean{M_i}\mean{M_j}.$ Since 
the linear part is compatible with convex combinations, the 
linear part on the right hand side of 
Eqs.~(\ref{concave}, \ref{krit1}) has to coincide with the 
linear part on the left hand side. 
The non-vanishing elements of $\gamma$ origin from the linear 
part only, so the missing $(\kappa_B)_{i,i}$ are
$(\kappa_B)_{i,i} = 1/d_B - \sum_k p_k \mean{G_i}^2_{\ketbra{b_k}},$
hence $(\kappa_B)_{i,i} \leq 1/d_B$.
So, if $\vr$ in the FNF is separable, then
\bea
	\sum_i \xi_i &\leq& 
	{d_A d_B} 
	\big(1-{1}/{d_A} +(d_A^2-1)/{d_B}
	 \label{stronger}
	\\
	&+&\min \{0; -(d_B-1) + (d_B^2-d_A^2)/{d_B} \}
	\big)/2.
	\nonumber
\eea
The possibility of taking zero in the minimization comes 
from the fact that one may also omit the summation over 
$(\kappa_B)_{i,i}.$ Especially if $d_A \ll d_B$ this yields
better bounds.

It is instructive to compare this with the CCNR criterion 
which requires for separable states in the FNF
$\sum_i \xi_i \leq d_A d_B- (d_A d_B)^{1/2},$
and the {dV-criterion} \cite{vicente}, requiring
$\sum_i \xi_i \leq \bigl({d_A d_B (d_A-1)(d_B-1)}\bigr)^{1/2}.$
For $d_A=d_B$ all three criteria coincide. One can directly 
see that for $d_A < d_B$ Eq.~(\ref{stronger}) is 
stronger than the CCNR criterion, but the dV-criterion is 
also stronger than CCNR. If $d_B -d_A$ is small, the 
dV-criterion is slightly better than  
Eq.~(\ref{stronger}), for  $d_A \ll d_B,$ however,  
Eq.~(\ref{stronger}) is drastically better than the CCNR and dV-criterion. 

{\it Connection with the LURs. --}
LURs allow detection of entanglement in the following way \cite{lurs}:
One takes observables $\{\hat A_k\}$ and $\{\hat B_k\}$ 
on Alice's (resp.\ Bob's) space and 
computes two bounds $U_A, U_B$ such that 
$\sum_k \delta^2(\hat A_k) \geq U_A  $,
$\sum_k \delta^2(\hat B_k) \geq U_B$. 
Then for separable states
$
\sum\nolimits_k 
\delta^2(\hat A_k \otimes \eins +\eins \otimes \hat B_k)
\geq U_A + U_B 
$
holds. Physically,  LURs show that separable 
states inherit the uncertainty relations from their 
reduced states, which is not the case for entangled states.
For a given state $\vr$, however, it is 
usually not clear which $\{\hat A_k\}$ and $\{\hat B_k\}$ 
are suitable to detect its entanglement.
We can formulate: 

\noindent
{\bf Proposition 7 (Connection to LURs).} 
{\it A state $\vr$ violates the CMC iff it can be detected by a LUR.}

{\it Proof.} The proof is given in the Appendix. 
\qed

This equivalence has two consequences:
Results concerning LURs can be transferred 
to the CMC. E.g.,
Ref.~\cite{ccn} gives now an alternative proof that the CMC 
is stronger than the CCNR criterion.
Also, our results concerning the CMC allow to gain new insights
in LURs:

\noindent
{\bf Corollary 8 (LURs for two qubits).} 
{\it 
There exist entangled two-qubit states which
can not be detected by a LUR.}

Examples of such states can be found numerically \cite{numerics},
showing that the filtering indeed improves the CMC.

{\it Examples and extensions. --}
Firstly, we take the $3\times 3$ 
bound entangled states arising from an {\it unextendible product 
basis}, mixed with white noise \cite{upb}. These states 
$\vr_{\rm UPB}(p)$ are detected by Proposition 6 for 
	$p \geq 0.8723$
while the best known positive map 
detects them only for $p> 0.8744$ \cite{ccn}. 
Secondly, we checked randomly 
generated $3\times 3$ {\it chessboard states} \cite{brussperes}.
The CCNR criterion detected 18.1 \% of them, Proposition 4 
detected 19.1 \% and Proposition 6 detected 97.8\%. 

Finally, let us note that the theory developed in this paper may be 
complemented by considering non-symmetric CMs \cite{Prep}. 
That is, one may define 
$\gamma^A_{i,j} = \mean{M_i M_j} - \mean{M_i}\mean{M_j}.$ For such 
matrices (which are now hermitian, but still positive) one can 
directly derive a statement corresponding to Proposition 1. This 
criterion implies then the CMC as given for a symmetric $\gamma.$

{\it Conclusion. --} We have proposed to investigate the 
separability of finite dimensional quantum states using the CM 
for certain observables. We have demonstrated that this approach
reveals the entanglement of many states and can lead to new 
insights into already existing criteria. It hence provides a further
systematic framework of studying entanglement for 
composite quantum systems. 

We thank H.J.\ Briegel, J.I.\ Cirac, R. Horodecki, M.\ van den Nest,
G. T\'oth and an anonymous referee for discussions and 
the FWF, the EU (OLAQUI, SCALA, QICS, QAP), 
the QIP-IRC, Microsoft Research, the DFG 
(SFB 407), and the EURYI  for support.

{\it Appendix. --} Here, we prove Proposition 7.
Note that the CM can be used to compute variances: if $N=\sum_k \nu_k M_k$
is a linear combination of the $M_k$ then 
$\delta^2(N)=\sum_{i,j}\nu_i \gamma(M_k)_{i,j} \nu_j 
=\bra{\nu}\gamma(M_k)\ket{\nu}$. 
If $\vr$ violates the LURs we can find $\hat A_k$ 
and $\hat B_k$ as above. We write 
$\hat A_k = \sum_l \alpha^{(k)}_{l} A_l$ 
and 
$\hat B_k = \sum_l \beta^{(k)}_{l} B_l$  with $A_k$ and $B_k$ 
as in the definition of $\gamma$ in Eq.~(\ref{gamdef}),
leading to 
$\delta^2(\hat A_k \otimes \eins +\eins \otimes \hat B_k)
=
\bra{\alpha^{(k)} \oplus \beta^{(k)}} 
\gamma 
\ket{\alpha^{(k)} \oplus \beta^{(k)}}.
$ 
For the $\kappa_A \oplus \kappa_B$ we have 
$\kappa_A \oplus \kappa_B := 
\sum_k p_k \gamma(\ketbra{a_k}) \oplus \gamma(\ketbra{b_k}),$ 
hence
$
\bra{\alpha^{(k)} \oplus \beta^{(k)}} 
\kappa_A \oplus \kappa_B
\ket{\alpha^{(k)} \oplus \beta^{(k)}} 
=
\sum_l p_l 
( \delta^2(\hat A_k)_{\ketbra{a_l}} + \delta^2(\hat B_k)_{\ketbra{a_l}}).
$
If the CMC were fulfilled, summing over $k$ would yield
$
\sum_k \delta^2  (\hat A_k \otimes \eins +\eins \otimes \hat B_k)
\geq 
\sum_l p_l 
\big( \sum_k \delta^2(\hat A_k)_{\ketbra{a_k}} + 
\sum_k \delta^2(\hat B_k)_{\ketbra{b_k}}\big)
\geq 
\min_{\ketbra{a}} 
(\sum_k\delta^2(\hat A_k)_{\ketbra{a}}) 
+ 
\min_{\ketbra{b}} 
(\sum_k\delta^2(\hat B_k)_{\ketbra{b}})
\geq U_A + U_B$,
which contradicts the violation of the 
LURs.

Conversely, let us define $X$ as 
the set of all matrices which can be written as 
$\kappa_A \oplus \kappa_B + P$ with some 
$\kappa_A$ and $\kappa_B$ as in Observation 1 and a positive $P$. 
In other words, the CMC states that for separable states
$\gamma \in X.$ Geometrically, $X$ is a closed convex 
cone. According to a Corollary of the Hahn-Banach Theorem 
for each $\gamma \not \in X$ there must be a 
symmetric matrix $W$ and a number $C$ such that 
$\tr[W\gamma]<C$ while $\tr[W x]>C$ for all $x \in X.$ 
Since $\tr[WP]\geq 0$ for all $P \geq 0,$ we have 
$W \geq 0.$ Now we use the spectral decomposition 
and  write 
$W = \sum_k \lambda_k \ketbra{\psi_k} = 
\sum_k \lambda_k \ketbra{\alpha^{(k)} \oplus \beta^{(k)}}.$ 
Defining 
$
\hat A_k = \sqrt{\lambda_k} \sum_l \alpha^{(k)}_l A_l, 
$
and
$
\hat B_k = \sqrt{\lambda_k} \sum_l \beta^{(k)}_l B_l, 
$ 
we have for $\vr$ that
$
\tr[W\gamma] = 
\sum_k \delta^2(\hat A_k \otimes \eins +\eins \otimes \hat B_k) < C$.

Furthermore, we have  that all  $\kappa_A \oplus \kappa_B \in X$ and 
even more, we have that all $\gamma_A \oplus \gamma_B \in X$. 
Hence,  for a product state
$\vr=\vr^A \otimes \vr^B$ we have
$
C < \tr[W(\gamma_A \oplus \gamma_B)]=  
(\sum_k \delta^2(\hat A_k)_{\vr^A} +\sum_k\delta^2(\hat B_k)_{\vr^B})
$
which implies that
$
C < \min_{\vr^A}
(\sum_k\delta^2(\hat A_k)_{\vr^A})
+
\min_{\vr^B}
(\sum_k\delta^2(\hat B_k)_{\vr^B}) 
=: U_A + U_B
\nonumber
$
leading to a violation of the LURs.
$\qed$


\begin{thebibliography}{99}
\bibitem{CVC}
	Note that the CMC bears great similarity with a criterion for 	
	continuous 
	variables \cite{wernerwolf,PhilippJens} and 
	has been mentioned, but not 
	further investigated in  Ref.~\cite{OtfriedsPRL}.
\bibitem{kapparemark}
        $\vr=\ketbra{1}$ and $\{M_k\}$ as in
        Ref.~\cite{ccn} is the simplest choice.
\bibitem{numerics}
	We have investigated randomly generated entangled two-qubit
	states, and checked the CMC via Proposition 2.
\bibitem{Remarkable}
	It is remarkable
	that the CCNR in fact follows from the theory, as
	Eq.~(\ref{eval1}) represents a single quadratic expression in 
	the elements of the density matrix.
\bibitem{remark}
        For that, one minimizes the function
        $
        f(\rho_A, \rho_B ) = 
        {\tr[\rho (\rho_A \otimes \rho_B)]}/(
        {\det(\rho_A)^{1/d_A}\det(\rho_B)^{1/d_B}}
        )$,
        takes the optimal $\rho_A, \rho_B$ and 
        defines $F_A^\dagger F_A =\rho_A$ and 
        $F_B^\dagger F_B =\rho_B.$
\bibitem{rankremark} For the two-qubit case, one uses a filtering as in
        Ref.\ \cite{frank}. In some cases, one will not directly arrive
        at the FNF, then one can use a further filter which brings $\vr$ 
        arbitrarily close to a maximally entangled state.
\end{thebibliography}
\end{document}